\documentclass[aps,12pt]{revtex4}
\usepackage{amsmath}
\def\[{\left\lbrack}
\def\]{\right\rbrack}

\def\({\left(}
\def\){\right)}
\newcommand{\be}{\begin{equation}}
\newcommand{\ee}{\end{equation}}
\newcommand{\ea}{\end{eqnarray}}
\newcommand{\ba}{\begin{eqnarray}}

\begin{document}

\title{Noncommutative Metafluid Dynamics}

\author{A.C.R. Mendes, C. Neves, W. Oliveira and F.I. Takakura }
\thanks{\noindent e-mail:albert@fisica.ufjf.br, cneves@fisica.ufjf.br,\\ wilson@fisica.ufjf.br, takakura@fisica.ufjf.br}
\affiliation{Departamento de F\'{\i}sica, Universidade Federal de
Juiz de Fora, 36036-330, Juiz de Fora, MG, Brasil}

\begin{abstract}
In this paper we define a noncommutative (NC) Metafluid Dynamics \cite{Marmanis}. We applied the Dirac's quantization to the Metafluid Dynamics on  NC spaces. First class constraints were found which are the same obtained in \cite{BJP}. The gauge covariant quantization of the non-linear equations of  fields on noncommutative spaces were studied. We have found the extended Hamiltonian which leads to equations of motion in the gauge covariant form. In addition, we show that a particular transformation \cite{Djemai} on the  usual classical phase space (CPS) leads to the same results as of the $\star$-deformation with $\nu=0$. Besides, we will shown that an additional term is introduced into the dissipative force due the NC geometry. This is an interesting feature due to the NC nature induced into model.
\end{abstract}

\maketitle

\section{Introduction}\label{intr}
The understanding of hydrodynamic turbulence is an important problem for nature
science, from both, theoretical and experimental point of view,
and has been investigated intensively \cite{Landau,Monin-Yagolm,Frisch} over the
last century, but a deep and fully comprehension of the problem remains obscure.
Over the last years, the investigation of turbulent hydrodynamics has experienced a
revival since turbulence has became a very fruitful
research field for theoreticians, that study the analogies between turbulence and
field theory, critical phenomena and condensed matter
physics
\cite{Bramwell-Holdsworth-Pinton,L'vov,Periwal,Polyakov,Gurarie,Eyink-Goldenfeld,Nelkin,deGennes,Rose-Sulem},
renewing the optimism to solve the turbulence problem.
A new approach for investigation of the fluid turbulence was proposed recently \cite{Marmanis}. The method named the metafluid dynamics, is  based on the use of analogy between Maxwell
electromagnetism and turbulent hydrodynamics, and describes the dynamical behavior of
average flow quantities in incompressible fluid flows
with high Reynolds numbers exactly as is done to obtain the macroscopic
electromagnetic fields \cite{Jackson}. In this method, the equations of motion describing the behavior of the hydrodynamic turbulence, after the average process (see more details Ref.\cite{Marmanis}),  are written as
\ba\label{1}
{\nabla .\vec  \omega} &=& 0,\nonumber \\
{{\partial } \over {\partial t}}\vec \omega &=& -{\nabla} \times {\vec l} + \nu {\nabla }^{2}{\vec \omega}, \nonumber \\ 
\nabla .\vec l &=& n({\vec x}, t), \\
{{\partial } \over {\partial t}}\vec l &=& \nabla \times \vec \omega - 
{\vec j}({\vec x}, t) + \nu {\nabla }n({\vec x}, t) - \nu {\nabla }^{2}{\vec l },\nonumber 
\ea
where $\vec w=\nabla \times \vec u$ is the vorticity, $\vec l=\vec \omega \times \vec u=-\frac{\partial \vec u}{\partial t} -\nabla \phi +\nu \nabla^2 \vec u$ is the Lamb vector, while  $\vec j$ is the turbulent current and $n$ is the turbulent charge, given by 
\be\label{2}
n(\vec x,t) =- \nabla^2 \phi .
\ee
In (\ref{2}) the Bernoulli energy function $\phi(\vec x,t)$ has the expression 
\be\label{2.1}
\phi(\vec x,t) =\frac{p}{\rho} + \frac{ u^2}{2},
\ee
where $p(\vec x,t)$ is the pressure, $\rho$ is the density and $\vec u(\vec x,t)$ represents the velocity field.

In recent work \cite{BJP}, we extended the Marmanis analogy in order to propose an appropriate Lagrangian governing the dynamics of incompresible fluid flows, in terms of velocity field and the Bernoulli energy function (named ``potentials of theory"), as
\begin{eqnarray} 
\label{3} 
{\cal L} = {1 \over 2} \left( -{\nabla }\phi -{{\partial } \over {\partial t}}\vec u
+ {\nu }{\nabla } ^{2}{\vec u} \right) ^{2} -{1 \over 2} ({\nabla } \times
{\vec u}) ^{2}+  {\vec j} .{\vec u}
-\phi n(\vec x,t)  - {\nu }{\vec u} .{\nabla }n(\vec x,t). 
\end{eqnarray}
It is  easy to see that this Lagrangian density gives us the equations of motion (\ref{1}) for viscous fluid \cite{Landau}. In \cite{BJP}, the theory was analyzed for the first time as a constrained system from the symplectic point of view \cite{FJ} and a hidden gauge symmetry was reported. 

In the case where the viscosity can be despised, as in the inertial range \cite{Frisch}, the Lagrangian density (\ref{3}) can be written as
\be\label{4}
{\cal L} =-\frac{1}{4}F_{\mu\nu} F^{\mu\nu} -J_{\mu}V^{\mu} ,
\ee
where the field strength tensor is
\be\label{5}
F_{\mu\nu} =\partial_\mu V_\nu -\partial_\nu V_\mu,
\ee
$V_\mu =(\vec u, \phi )$ is the vector-potential of the Lamb and vorticity field and $J_\mu =(\vec j,n)$. Its opportune to comment here that the Metafluid Dynamics was investigated within Hamilton-Jacobi formalism \cite{Dumitru}.

In the present paper, we will apply the Dirac's quantization to the Metafluid Dynamics on (NC) spaces from the $\star$-deformation of the  (CPS), where  the extended Hamiltonian leading to gauge covariant equations of motion is derived. In addition, we show that a particular transformation on the  usual CPS leads to the same results as of the $\star$-deformation on CPS.

This paper is organized as follows. In section \ref{field} the general Dirac's procedure of quantization of $\star$-deformed metafluid dynamics is considered. The extended Hamiltonian leading to gauge covariant equations of motion is derived. In section \ref{quan} the gauge fixing approach on the basis of Dirac's brackets is studied. In section \ref{trans}, a particular transformation on the commutative configuration space is considered, introducing noncommutativity on the configuration space. Section \ref{concl} is devoted to the discussion.

\section{The Noncommutative Metafluid Dynamics}\label{field}

The Lagrangian density (\ref{4}) on NC space is given by

\ba\label{01}
{\cal L} &=&-\frac{1}{4}F_{\mu\nu}\star F^{\mu\nu} -J_{\mu}\star V^{\mu} \\
&=& -\frac{1}{4}\hat F_{\mu\nu}^2 -\hat J_{\mu} \hat V^{\mu}
\ea
where noncommutative strength  $\hat F_{\mu\nu}$ reads
\be\label{02}
\hat F_{\mu\nu} = \partial_{\mu} V_{\nu} -\partial_\nu V_\mu -i\( V_{\mu} \star V_{\nu} -V_{\nu}\star V_{\mu} \) .
\ee
Here the star product ($\star$)  between NC quantities is defined as usual:
\be\label{02.1}
\left. V_\mu (\vec x) \star V_\nu (\vec y) =\exp \( \frac{i}{2} \theta^{ij} \partial_i^x \partial_j^y \)V_\mu(\vec x)V_\nu (\vec y) \right|_{x=y}.
\ee
The Seiberg-Witten expansion to the first order in $\theta_{\mu\nu}$ \cite{Bichl} gives
\begin{subequations}\label{03}
\begin{gather}
\hat A_{\mu} = A_{\mu} -\frac{1}{2}\theta_{\alpha\beta}A_{\alpha}\(\partial_{\beta}A_{\mu} +F_{\beta\mu}\), \label{3a}\\
\hat F_{\mu\nu} = F_{\mu\nu} + \theta_{\alpha\beta}F_{\mu\alpha}F_{\nu\beta} -\theta_{\alpha\beta}A_{\alpha}\partial_{\beta}F_{\mu\nu}.\label{3b}
\end{gather}
\end{subequations}
On applying this map, the Lagrangian density (\ref{01}) is written in terms of ordinary fields in the order of ${\cal O}(\theta^2)$ , as
\be\label{04}
{\cal L} =-\frac{1}{4}F_{\mu\nu}^2 +\frac{1}{2}F_{\mu\nu} \theta_{\alpha\beta}V_{\alpha}\partial_{\beta} F_{\mu\nu} -\frac{1}{2}F_{\mu\nu}\theta_{\alpha\beta} F_{\mu\alpha}F_{\nu\beta}+ J_\mu V_\mu +{\cal O}(\theta^2),
\ee
where we took into consideration that the term $J_{\mu}\star V_{\mu}$ in the Lagrangian on  NC spaces coincides within four-divergences with $J_{\mu}V_{\mu}$. The Lagrangian density (\ref{04}) can also be cast in the form of
\be\label{05}
{\cal L} = \frac{1}{2}( \vec l^2 - \vec \omega^2 )\[ 1+ ( \theta .\vec\omega )\] -( \theta .\vec l )( \vec l . \vec \omega) + V_{\mu}J_{\mu}+ {\cal O}(\theta^2 ) ,
\ee
where $\theta_i = (1/2)\epsilon_{ijk}\theta_{jk}$, $\theta_{i4}=0$. Using the Euler-Lagrange equations
\be\label{06}
\partial_{\mu} \frac{\partial{\cal L}}{\partial (\partial_{\mu}V_{\nu} )} -\frac{\partial{\cal L}}{\partial V_{\nu}} =0,
\ee
we obtain from (\ref{04}) the following field equations
\ba\label{07}
&&\partial_\mu F_{\mu\nu} + \frac{1}{2} \theta_{\alpha\beta}\partial_\mu \( F_{\mu\nu} F_{\alpha\beta}\) +\frac{1}{4}\theta_{\mu\nu}\partial_\mu \(F_{\alpha\beta}^2 \)\nonumber \\ &&- \theta_{\nu\beta}\partial_\mu \(F_{\alpha\beta}F_{\mu\nu} \) + \theta_{\mu\beta} \partial_\mu \(F_{\alpha\beta}F_{\nu\alpha} \) -\theta_{\alpha\beta}\partial_{\mu} \(F_{\mu\alpha}F_{\alpha\beta} \) =J_\nu
\ea
The non-linear equations (\ref{07}) may be cast as follows:
\begin{subequations}\label{08}
\begin{gather}
\frac{\partial}{\partial t}\vec L -\nabla \times \vec W = -\vec j(\vec x,t) ,\label{8a}\\
\nabla . \vec L =n(\vec x,t) \label{8b}
\end{gather}
\end{subequations}
where the fields $\vec L$ and $\vec W$ are
\be\label{09}
\vec{ L}= \vec l + \vec \ell , \;\; \vec \ell= ( \theta .\vec \omega )\vec l -(\theta .\vec l \;) \vec \omega -(\vec l.\vec \omega )\theta ,
\ee
\be\label{10}
\vec W = \vec w + \vec \varpi , \;\; \vec \varpi = (\theta .\vec \omega )\vec \omega + (\theta .\vec l \;)\vec l -\frac{1}{2} ({\vec l}\;^2 -\vec \omega^2 ).
\ee
The equations (\ref{8a}) and (\ref{8b}) are the two last equations for the metafluid dynamics, (\ref{1}), with $\nu=0$.

The second pair of equations, which is the consequence of (\ref{5}) is 
\be\label{11}
\partial_\mu \tilde F_{\mu\nu} =0,
\ee
where the dual tensor being $\tilde F_{\mu\nu}=\frac{1}{2}\epsilon_{\mu\nu\alpha\beta}F_{\alpha\beta}$, $\epsilon_{\mu\nu\alpha\beta}$ is a Levy-Civita antisymmetric tensor. So, (\ref{11}) is rewritten as
\begin{subequations}\label{12}
\begin{gather}
\frac{\partial \vec w}{\partial t} +\nabla \times \vec l =0,\label{12a}\\
\nabla .\vec w =0.\label{12b}
\end{gather}
\end{subequations}
that are the  two first equations for the metafluid dynamics, (\ref{1}), with $\nu=0$.

Now we apply Dirac's formalism of gauge covariant quantization to the Lagrangian (\ref{04}) which leads to non-linear field equations. According to the Dirac formalism, we find from (\ref{04}), with the accuracy of ${\cal O}(\theta^2)$, the following momenta
\begin{subequations}\label{14}
\begin{align}
\pi_i &= \frac{\partial {\cal L}}{\partial (\partial_0 V_i )} =-l_i [1+(\theta.\vec \omega)]+(\theta .\vec l \;)\omega_i +(\vec l .\vec w)\theta_i ,\label{14a}\\
\pi_0 &=\frac{\partial {\cal L}}{\partial (\partial_0 V_0 )} =0.\label{14b}
\end{align}
\end{subequations}
We observe that $\pi_0$ is a primary constraint, see Ref.\cite{Dirac} that we denote by $\chi_1$,
\be\label{15}
\chi_1 (\vec x) =\pi_0 \approx 0.
\ee
From equations (\ref{09}) and (\ref{14a}), we come to the equality $\pi_i =-L_i$. Then, using the known Poisson brackets between coordinates $V_i (\vec x)$ and momentum $\pi_i$, we arrive at
\be\label{16} 
\{ V_i (\vec x,t),L_i (\vec y, t) \} =-\delta_{ij} \delta (\vec x -\vec y ).
\ee
From (\ref{16}) it is easy to find the Poisson bracket between the vorticity field $\vec \omega$ and the field $\vec L$:
\be\label{17}
\{\omega_i (\vec x ,t), L_j (\vec y,t) \} =\epsilon_{ijk}\partial_k \delta (\vec x -\vec y).
\ee
The density of the Hamiltonian found from the relation ${\cal H} =\pi_\mu \partial_0 V_\mu -{\cal L}$, with the help of (\ref{05}) and (\ref{14a}), is given by
\be\label{18}
{\cal H}=\frac{1}{2}(\vec l\;^2 + \vec \omega^2 )[1+ (\theta .\vec \omega )] -(\theta .\vec l\; )(\vec l .\vec \omega )-\vec \pi . \nabla V_0 + V^\mu J_\mu .
\ee

The primary constraint (\ref{15}) should be a constant of motion and, therefore, we have the condition
\ba\label{19}
\partial_0 \pi_0 (\vec x) =\{\pi_0 ,H \} &=&\int d^3 y \{\pi_0 (\vec x) ,{\cal H}(\vec y) \} \nonumber \\
&=& -\partial_i \pi_i -n(\vec x, t) =0,
\ea
where
\be\label{20}
H=\int d^3 y \;{\cal H}
\ee
is the Hamiltonian. The (\ref{19}) guarantees that the primary constraint (\ref{15}) is conserved, i.e., has no time dependence. Hence, the secondary constraint, found from (\ref{19}), is
\be\label{21}
\chi_2 (\vec x) = \partial_i \pi_i + n(\vec x, t) \approx 0.
\ee 
Note  that, using equality $\pi_i =-L_i$, it is easy to see that the secondary constraint (\ref{21}) is simply (\ref{8b}).
The time evolution of the secondary constraint is
\be\label{22}
\partial_0 \chi_2 (\vec x) =\{ \chi_2 (\vec x), H\} = -\partial_i J_i (\vec x).
\ee
But, from (\ref{1}), we get
\be\label{23}
\partial_i J_i = -\nabla^2 \frac{\partial }{\partial t}\phi,
\ee
where, in the inertial range, the Bernoulli energy function $\phi$ is constant \cite{Marmanis}, so
\be\label{24}
\partial_0 \chi_2 (\vec x) =\{ \chi_2 (\vec x), H\} = -\partial_i J_i (\vec x)\equiv 0.
\ee
This shows that there is not additional constraint. The Poisson brackets between  $\varphi_1$ and $\varphi_2$,respectively primary and secondary constraints, vanishes, $\{\varphi_1 ,\varphi_2 \}=0$. Thus, all constraints here are first class, and there are no second class constraints. According to the general Dirac method, to acquire the total density of Hamiltonian, we add to (\ref{18}) Lagrange multiplier terms $\lambda (\vec x)\pi_0 $, $\beta(\vec x)(\partial_i \pi_i + n)$, where $\lambda(\vec x)$ and $\beta(\vec x)$ are auxiliary variables which have no physical meaning, and are connected with gauge degrees of freedom. Then, we get
\ba\label{25}
{\cal H}_T =&&\frac{1}{2}(\vec l\;^2 + \vec \omega^2 )[1+ (\theta .\vec \omega )] -(\theta .\vec l\; )(\vec l .\vec \omega )+ V^\mu J_\mu\nonumber\\
&&-\vec \pi . \nabla V_0 +\lambda (\vec x)\pi_0 + \beta(\vec x)(\nabla . \vec \pi + n).
\ea

To obtain equations of motion we have to express the density of Hamiltonian (\ref{25}) in  terms of the fields, $V_\mu$, and momenta, $\pi_\mu$. For this, using (\ref{14a}), we find, with the accuracy of ${\cal O}(\theta^2)$, the Lamb field is
\be\label{26}
l_i =-\pi_i [1- (\theta .\vec \omega)]-(\theta . \vec \pi )\omega_i -(\vec \pi .\vec \omega )\theta_i,
\ee                                                                                                   
together with (\ref{26}) and the equality $L_i =-\pi_i$, the total density of Hamiltonian (\ref{25}) takes the form
\ba\label{27}
{\cal H}_T =&&\frac{1}{2}( \vec \pi ^2 +\vec \omega ^2 ) + (\theta .\vec \omega)\frac{1}{2} (\vec \omega ^2 -\vec \pi ^2 ) +(\theta .\vec \pi )(\vec \pi . \vec\omega ) + \nonumber\\
&&+[V_0 + \beta(\vec x) ](\nabla . \vec  \pi +n ) +\lambda(\vec x) \pi_0   - V_i J_i .
\ea                                                  
The total   Hamiltonian density allows us to obtain the time evolution of the fields. With the help of the Hamiltonian equations we find
\ba\label{28}
\partial_0 V_i &=&\{V_i , H \} \nonumber\\
&=& \pi_i [1- (\theta .\vec \omega )] + \theta_i (\vec \pi .\vec \omega ) + (\theta .\vec \pi )\omega_i -\partial_i V_0 -\partial_i \beta(x);
\ea
\ba\label{29}
\partial_0 \pi_i &=& \{ \pi_i ,H \} \nonumber\\
&=&\partial_n ([\partial_n V_i -\partial_i V_n ][(\theta .\vec \omega)+1])+\epsilon_{ilm}\partial_m \(\theta_l \frac{\vec \omega^2 -\vec \pi^2}{2} +\pi_l (\vec \pi .\theta ) \)-J_i ;
\ea
\be\label{30}
\partial_0 V_0 =\{ V_0 ,H \} = \frac{\delta H}{\delta \pi_0 } =\lambda (x), \;\;\;\; \partial_0 \pi_0 =\{ \pi_0 ,H\}=-\frac{\delta H}{\delta V_0 }= -\partial_i \pi_i -n(\vec x,t).
\ee
Equation (\ref{29}) coincides with the first equation in (\ref{8a}) taking into consideration the definition (\ref{09}) and (\ref{10}), and (\ref{28}) is nothing but the gauge covariant form of (\ref{26}). (\ref{30}) represents the time evolution of non-physical fields.

\section{Quantization of second-class constraints}\label{quan}

In this section, we consider gauge fixing approach.
With  help of the gauge freedom, decribed by two functions, $\lambda(x)$, $\beta(x)$, we can impose new constraints as follows \cite{BJP}:
\be\label{31}
\chi_3 (\vec x)= V_0 \approx 0,\;\;\; \chi_4 (\vec x) =\partial_i V_i \approx 0 .
\ee
The Coulomb gauge (\ref{31}) does not violate the equations of motion. After fixing two components of vector-potential $V_\mu$ in accordance with (\ref{31}), the first class constraints (\ref{15}) and (\ref{21}) become second class constraints. Indeed, the non-zero Poisson brackets of the functions $\chi_1$, (\ref{15}), $\chi_2$, (\ref{21}) and $\chi_3$, $\chi_4$, (\ref{31}), are
\ba\label{32}
\{ \chi_1 (\vec x),\chi_3 (\vec y)\} &=& -\delta(\vec x-\vec y),\nonumber\\
\{\chi_2 (\vec x),\chi_4 (\vec y) \} &=& -\partial_x^2 \delta(\vec x -\vec y).
\ea
Defining the matrix of Poisson brackets as
\be\label{33}
C_{ij} =\{ \chi_i (\vec x,t),\chi_j (\vec y,t)\},
\ee
so that the inverse matrix $C_{ij}^{-1}$ exists, we may introduce the Dirac brackets \cite{Dirac}
\ba\label{34}
\{ A(\vec x,t),B(\vec y,t) \}^* &=&\{ A(\vec x,t),B(\vec y,t)\} + \nonumber\\
&-&\int d^3 z d^3 s \;\; \{A(\vec x,t),\chi_i (\vec z,t)\} C_{ij}^{-1}\{ \chi_j (\vec s,t),B(\vec y,t)\}.
\ea
The inverse matrix $C_{ij}^{-1}$ obeys the equation
\be\label{35}
\int d^3 z\;\; C_{ik}(\vec x,\vec z)C_{kj}^{-1}(\vec z,\vec y) =\delta_{ij}\delta(\vec x-\vec y),
\ee
and is given by
\be\label{36}
C^{-1} =\left(
\begin{array}{cccc}
0 & 0 & \delta(\vec x-\vec y) & 0 \\
0 & 0 & 0 & \frac{1}{4\left| \vec x -\vec y \right| }\\
-\delta (\vec x-\vec y ) & 0 & 0 & 0 \\
0 & -\frac{1}{4\left| \vec x -\vec y \right| } & 0 & 0 
\end{array}\right).
\ee
Using the definition of Dirac brackets (\ref{34}) and (\ref{36}), and imposing the boundary condition that the fields vanish at infinity, we get
\begin{subequations}\label{37}
\begin{align}
\{ \pi_0(\vec x),V_0 (\vec y) \}^* &=\{\pi_0 (\vec x), V_i (\vec y) \}^* =\{ \pi_i (\vec x),V_0 (\vec y) \}^* =0, \label{37a}\\
\{ \pi_i (\vec x), V_j (\vec y)\}^* &= -\delta_{ij}\delta (\vec x-\vec y) +\frac{\partial^2}{\partial x_i \partial y_j}\frac{1}{4\pi \left| \vec x -\vec y \right|}, \label{37b}\\
\{ \pi_\mu (\vec x), \pi_\nu (\vec y) \}^* &=\{ V_\mu (\vec x), V_\nu (\vec y) \}^* =0.
\end{align}
\end{subequations}
Those Dirac bracktes are the same obtained in \cite{BJP} from the Faddeev-Jackiw approach.

Now, following the prescription \cite{Dirac}, we can set all second class constraints strongly to zero. Then the physical Hamiltonian of fully constrained $\star$-deformed theory becomes
\be\label{41}
H_{ph}=\int d^3 \vec x \; \left[\frac{1}{2}(\vec l\;^2 + \vec \omega^2 )[1+ (\theta .\vec \omega )] -(\theta .\vec l\; )(\vec l .\vec \omega ) - V_i J_i \right].
\ee
Equations of motion obtained from (\ref{41}) are given by
\be\label{42}
\partial_0 \vec V = \{\vec V, H_{ph} \}^* =
 \vec \pi [ 1- (\theta .\vec \omega) ]+ (\vec \pi . \vec \omega )\theta + (\theta . \vec \pi ) \vec \omega,
\ee
\be\label{43}
\partial_0 \vec \pi = \{\vec \pi , H_{ph}\}^* 
=-\nabla \times \(\vec \omega [ 1+(\theta .\vec \omega )] + \theta \frac{\vec \omega^2 -\vec \pi^2}{2} + \vec \pi (\vec \pi .\theta) \) + \vec J.
\ee
Now, using that $V_i =u_i$ (velocity field) and (\ref{09}), where $\pi_i =- L_i$, we have from (\ref{42}) that
\ba\label{44}
\partial_0 \vec u(\vec x) &=& -(\vec l + \vec \ell)[1-(\theta .\vec \omega)] +[-(\vec l + \vec \ell).\vec \omega ]\theta +[-(\vec l + \vec \ell).\theta ]\vec \omega \nonumber\\
&=& -\vec l +\vec \ell (\theta.\vec \omega)-(\vec \ell.\vec \omega )\theta -(\vec \ell .\theta )\vec \omega ,
\ea
or
\be\label{45}
\frac{\partial}{\partial t}\vec u = -(\vec \omega \times \vec u ) -\nabla \phi + {\cal O}(\theta^2),
\ee
where $\vec \omega \times \vec u =\vec l$. The equation (\ref{45}) is the Euler equation in the order of  $\theta^2$. In the quantum theory, with the presence of second class constraints, we have to replace Dirac's bracket by the quantum commutator according to the prescription $\{ . , . \}^* \rightarrow -i[ . , . ]$. Only transverse components of the vector potential $V_\mu$ are physical degrees of freedom, and they remain in the theory.
\section{A particular transformation}\label{trans}

Let us consider a general linear transformation on the usual CPS variables, given by
\be\label{46}
\left\{
\begin{array}{l}
u^\prime_i = a_{ik}u_k +b_{ik} \pi_k \nonumber\\
\pi^\prime_i = c_{ik} u_k + d_{ik} \pi_k .
\end{array}\right.
\ee
At this stage, we remark that for the following particular choice of the matrix parameters $a,b,c$ and $d$
\be\label{47}
a=d=1, \; c=0, \; b=-\frac{1}{2}\theta,
\ee
the new classical variables $u^\prime_i$ and $\pi^\prime_i$ are
\be\label{48}
\left(
\begin{array}{c}
 u^\prime_i \\
\pi^\prime_i 
\end{array}
\right)=
\left(
\begin{array}{cc}
\delta_{ik} & -\frac{1}{2}\theta_{ik}\\
0 & \delta_{ik}
\end{array}\right)
\left(
\begin{array}{c}
u_k \\ \pi_k
\end{array}
\right)
\ee
or, in another manner
\be\label{49}
\left\{
\begin{array}{l}
\vec u^\prime =\vec u -\frac{1}{2}\vec \pi \times \theta \\
\vec \pi^\prime =\vec \pi
\end{array}\right. .
\ee
These new fields satisfy the relations above,
\be\label{50}
\{ u_i^\prime , u_j^\prime \} =\theta_{ij}, \;\; \{u_i^\prime , \pi_j^\prime \} =\delta _{ij}, \;\; \{\pi_i^\prime ,\pi_j^\prime \}=0.
\ee
Note that, those transformations imply  the presence of a noncommutativity on the configuration space.

In terms of the new fields $u_i^\prime$, the Lagrangian density, (\ref{3}), is given by
\be\label{51}
{\cal L}^\prime =\frac{1}{2} \( -\partial_i \phi -\frac{\partial}{\partial t}u_i^\prime + \nu \partial^2 u_i^\prime \)^2 -\frac{1}{2}\(\epsilon_{ijk}\partial_j u_k^\prime \) + u_i^\prime j_i -n(\vec x,t)\phi -\nu u_i^\prime \partial_i n(\vec x,t);
\ee
that can be represented as
\ba\label{52}
{\cal L}^\prime &=&\frac{1}{2}\( -\partial_i \phi -\frac{\partial}{\partial t}u_i +\frac{1}{2} \theta_{ij}\frac{\partial}{\partial t}\pi_{j} +\nu\partial^2 u_i -\frac{\nu}{2}\theta_{ij}\partial^2 \pi_{j} \)^2\nonumber\\ 
&-&\frac{1}{2} \( \epsilon_{ijk}\partial_j u_k -\frac{1}{2} \epsilon_{ilm}\theta_{nm}\partial_l \pi_m \)^2 \nonumber\\ &+& u_i j_i -\frac{1}{2}\theta_{ij} u_j j_i +n(\vec x,t)\phi -\nu u_i \partial_i n(\vec x,t) +\frac{\nu}{2}\theta_{ij}\pi_j \partial_i n,
\ea
where $\partial_i =\frac{\partial}{\partial x_i}$ and $\partial^2 =\frac{\partial^2}{\partial x_i^2}$. Applying Euler-Lagrange equations for the velocity field $\vec u$, one gets
\ba\label{53}
&&\frac{\partial}{\partial t}\( \partial_i  \phi +\frac{\partial}{\partial t}u_i -\nu \partial^2 u_i -\frac{1}{2}\theta_{ij}\frac{\partial}{\partial t}\pi_j +\frac{\nu}{2}\theta_{ij}\partial^2 \pi_j \)\nonumber\\
&+& \epsilon_{ils}\partial_l \( \epsilon_{sjk}\partial_j u_k -\frac{1}{2}\epsilon_{srn}\theta_{nm}\partial_r \pi_m \)+\nu \partial_i n(\vec x,t) -\( j_i + \frac{1}{2}\theta_{ij} j_j \)\nonumber\\
&+&\nu \partial^2 \( \partial_i  \phi +\frac{\partial}{\partial t}u_i -\nu \partial^2 u_i -\frac{1}{2}\theta_{ij}\frac{\partial}{\partial t}\pi_j +\frac{\nu}{2}\theta_{ij}\partial^{2} \pi_{j} \) =0.
\ea
Inserting the expression of $\vec l$ and $\vec \omega$, the (\ref{53}) may be cast as
\be\label{54}
\frac{\partial}{\partial t}\vec L =  \nabla \times  \vec W -\vec J +\nu \nabla n(\vec x,t) -\nu \nabla^2 \vec L,
\ee
where the Lamb $(\vec L)$ e vorticity $(\vec W)$ fields are given by
\be\label{55}
\vec L = \vec l + \vec \ell\; ,\;\; \vec \ell = \frac{1}{2} \frac{\partial}{\partial t}(\vec \pi \times \theta ) -\frac{\nu}{2} \partial^2 (\vec \pi \times \theta ),
\ee
\be\label{56}
\vec W = \vec \omega + \vec \varpi \; , \;\; \vec \varpi =-\frac{1}{2} \nabla \times (\vec \pi \times \theta)
\ee
and
\be\label{57}
\vec J = \vec j + \frac{1}{2}\vec j \times \theta.
\ee 
Now, applying the Euler-Lagrange equations for $\phi$, one gets
\be\label{58}
\partial_i \( \partial_i  \phi +\frac{\partial}{\partial t}u_i -\nu \partial^2 u_i -\frac{1}{2}\theta_{ij}\frac{\partial}{\partial t}\pi_j +\frac{\nu}{2}\theta_{ij}\partial^{2} \pi_{j} \) -n(\vec x,t) =0.
\ee
Using (\ref{55}), we get
\be\label{59}
\nabla . \vec L = n(\vec x,t).
\ee
The equations of motion (\ref{54}) and (\ref{59}) are the two last equations for the metafluid dynamics, (\ref{1}). The other two equations are obtained directly from $\vec \omega$ and $\vec l$ definitions. Taking the divergence of $\vec \omega$, we get the first of (\ref{1}) and, in order to get the second one, take the curl of $\vec l$.

We find from (\ref{52}), with the accuracy of ${\cal O}(\theta^2)$, the following momenta
\ba\label{60}
\pi_i =\frac{\partial {\cal L}^\prime}{\partial \(\frac{\partial u_i}{\partial t }\)}&=& \partial_i \phi +\frac{\partial}{\partial t}u_i -\nu \partial^2 u_i -\frac{1}{2}\theta_{ij}\frac{\partial}{\partial t}\pi_j +\frac{\nu}{2}\theta_{ij}\partial^2 \pi_j \nonumber \\
&=& -l_i -\frac{1}{2}\theta_{ij}\frac{\partial}{\partial t}\pi_j +\frac{\nu}{2}\theta_{ij}\partial^2 \pi_j ,
\ea
therefore, from (\ref{60}) and (\ref{55}), we come to the equality $\pi_i  = -L_i$.
Using (\ref{55}), we find that
\ba\label{60.1}
\pi_i &=& -l_i -\frac{1}{2}\theta_{ij}\frac{\partial}{\partial t}(-l_j -\ell_j ) +\frac{\nu}{2}\theta_{ij}\partial^2 (-l_j -\ell_j ) \nonumber\\
&=& -l_i +\frac{1}{2}\theta_{ij}\frac{\partial}{\partial t}l_j -\frac{\nu}{2}\theta_{ij}\partial^2 l_j + {\cal O}(\theta^2).
\ea

The density of the Hamiltonian found from the relation ${\cal H}^\prime =\pi_i^\prime \dot u_i^\prime - {\cal L}^\prime$, with the help of (\ref{51}) \cite{BJP}, is given by
\be\label{61}
{\cal H}^\prime  =\frac{1}{2} \pi_i^{\prime 2} -\pi_i^\prime \partial_i \phi + \nu \pi_i^\prime \partial^2 u_i^\prime -\frac{1}{2} ( \epsilon_{ijk}\partial_j u_k^\prime )^2 +\nu u_i^\prime \partial_i n(\vec x,t) -u_i^\prime j_i +\phi n(\vec x,t ).
\ee
Substituting (\ref{49}) in ${\cal H}^\prime $, we get
\ba\label{62}
{\cal H}^\prime &=& \frac{1}{2} \pi_i^2 -\pi_i \partial_i \phi +\nu \pi_i \partial^2 u_i -\frac{\nu}{2} \pi_i \theta_{ij} \partial^2 \pi_j +\nonumber\\
&-&\frac{1}{2} \( \epsilon_{ijk}\partial_j u_k -\frac{1}{2} \epsilon_{ijk}\theta_{kl}\partial_j \pi_l \)^2 +\nu u_i \partial_i n(\vec x,t) +\nonumber\\
&-&\frac{\nu}{2} \theta_{ik} \partial_i n(\vec x,t)\pi_k -u_i j_i +\frac{1}{2}\theta_{ij}j_i \pi_j + \phi n(\vec x,t).
\ea
In this case, the  noncommutative Hamiltonian's equations are
\ba\label{62.1}
\frac{\partial}{\partial t}\pi_i &=& \{ \pi_i , H^\prime  \} =\frac{\delta H^\prime }{\delta u_i }\nonumber\\
&=& -\nu\partial^2 \pi_i +\epsilon_{inm}\epsilon_{mjk}\partial_n \partial_j u_k - \frac{1}{2} \epsilon_{inm}\epsilon_{mjk}\theta_{kl} \partial_n \partial_j \pi_l -\nu \partial_i n(\vec x,t) + j_i ,
\ea
and
\ba\label{63}
\frac{\partial}{\partial t}u_i &=& \{ u_i ,H^\prime \} =\frac{\delta H^\prime}{\delta \pi_i}\nonumber\\
&=& \pi_i -\partial_i \phi +\nu \partial^2 u_i -\frac{\nu}{2}\theta_{ij}\partial^2 \pi_j -\frac{1}{2} \theta_{ij}\epsilon_{jkl}\epsilon_{lnm}\partial_k\partial_n u_m +\nonumber\\ &+&\frac{\nu}{2}\theta_{ij}\partial_j n(\vec x,t) +\frac{1}{2} \theta_{ij}j_i \;,
\ea
where
\be\label{63.1}
H^\prime  =\int d^3 \vec x \; {\cal H}^\prime 
\ee

Now, substituting (\ref{60.1}) in (\ref{63}), we have
\be\label{64}
\underbrace{\frac{\partial}{\partial t} \vec u +\nabla \phi -\nu\nabla^2 \vec u + \vec l}_{{\rm{NS \;\;equation}}} =\frac{1}{2} \( \frac{\partial}{\partial t}\vec l\) \times \theta -\frac{1}{2} \left[(\nabla \times \vec \omega) \times \theta \right]  -\frac{\nu}{2}\left[\nabla n(\vec x,t)\right]\times \theta +
\frac{1}{2} \vec j \times \theta +{\cal O}(\theta^2).
\ee 
Using the fourth equation in (\ref{1}), may be cast the (\ref{64}) on the usual configuration space as
\be\label{65}
\frac{\partial}{\partial t} \vec u +\nabla \phi -\nu\nabla^2 \vec u + \vec l =-\frac{\nu}{2}[ \nabla^2 \vec l ] \times \theta +{\cal O}(\theta^2).
\ee
The left side of this equation represents the usual expression of the Navier-Stokes equation of turbulent viscous fluid, in the noncommutative configuration space ($\theta =0$). The right side expresses a first correction to the dissipative force due to the viscosity depending on the presence of a noncommutativity on the configuration space ($\theta \neq 0$). Note that, when the viscosity is not present ($\nu=0$) the (\ref{65}) leads to the same results as from the $\star$-deformation on classical phase space. It is easy to see that the classical limit is recovered when the parameter $\theta \rightarrow 0$.

\section{Conclusion}\label{concl}

We have considered quantization of the Metafluid Dynamics on noncommutative spaces taking into consideration first class constraints as well as introducing second class constraints and the Dirac bracket. The procedure of Dirac's quantization here, on the basis of first class constraints and the Poisson bracket, is similar to the quantization obtained in \cite{BJP}. The difference is that field equations are nonlinear in the case of the $\star$-deformed theory. Afterward, by a particular transformation on the usual CPS leads to the same results as of the $\star$-deformation on CPS. We have shown that there is a correction term to the dissipative force, due to the viscosity,  that depends of the noncommutative parameter on the configuration space.


\begin{thebibliography} {99}

\bibitem{Marmanis} Haralambos Marmanis, Phys. Fluids {\bf 10} (1998) 1428; PhD
these: {\it Analogy between the Electromagnetic and Hydrodynamic Equations:
Application to Turbulence}, 2000, http://www.cfm.brown.edu/people/marmanis/.
\bibitem{Djemai} A.E.F. Djemaï and H. Smail, hep-th/0309006.
\bibitem{BJP} A.C.R. Mendes, C. Neves, W. Oliveira and F.I. Takakura, Braz. J. Phys. {\bf 33} (2003) 346.
\bibitem{Dumitru} Dumitru Baleanu, {\it Metafluid dynamics and Hamilton-Jacobi formalism}, hep-th/0412051.
\bibitem{Landau}L.D. Landau and E.M. Lifshits, {\it Fluid Mechanics} (Pergamon
Press, Oxford, 1980).
\bibitem{Monin-Yagolm} A.S. Monin and A.M. Yagolm, {\it Statistical Fluid Mechanics:
Mechanics of Turbulence} (The MIT Press, Cambridge, 1971).
\bibitem{Frisch} U. Frisch, {\it Turbulence: the Legacy of A.N. Kolmogorov},
Cambridge University Press, Cambridge (1995).
\bibitem{Bramwell-Holdsworth-Pinton} S.T. Bramwell, P.C.W. Holdsworth and J.F.
Pinton, Nature {\bf 396}, 552 (1998).
\bibitem{L'vov} V.S. L'vov, Nature {\bf 396} (1998) 519 .
\bibitem{Periwal} V. Periwal, cond-mat/9602123 (1996).
\bibitem{Polyakov} A.M. Polyakov, Nucl. Phys. {\bf B396} (1993) 367.
\bibitem{Gurarie} V. Gurarie, {\it Field Theory and the Phenomenon of Turbulence},
(Proc. Recent progress in statistical mechanics and quantum field theory, Los
Angeles, USA)(1994).
\bibitem{Eyink-Goldenfeld} G. Eyink and N. Goldenfeld, Phys. Rev. {\bf E50}
(1994)  4679. 
\bibitem{Nelkin} M. Nelkin, Phys. Rev. {\bf A9} (1974)  388 .
\bibitem{deGennes} P.G. De Gennes, {\it Fluctuation, Instability and Phase
Transition}, (Proc. NATO Adv. Study Inst., Geilo, Norway). T. Riste, ed. (Noordhoff,
Leiden), series B (1975), p1.
\bibitem{Rose-Sulem} H.A. Rose and P.L. Sulem, J. de Phys. {\bf 39} (1978) 441.
\bibitem{Jackson} J.D. Jackson, {\it Classical Electrodynamics} (J. Willey, New
York, 1983)
\bibitem{FJ} L. Faddeev and R. Jackiw, Phys. Rev. Lett. {\bf 60} (1988)  1692;\\
J. Barcelos Neto and C. Wotzasek, Mod. Phys. Lett. {\bf A7} (1992) 1737; Int. J.
Mod. Phys. {\bf A7} (1992) 4981 .
\bibitem{Bichl} A. Bichl, J. Grimstrup, L. Popp, M. Schweda and R. Wulkenhaar, hep-th/0102044.
\bibitem{Dirac} P.A.M. Dirac, Lectures on Quantum Mechanics (Yeshiva University, New York, 1964).



\end{thebibliography}
\end{document}